\documentclass[preprint,11pt]{elsarticle}

\usepackage{fullpage,amsmath,amssymb,graphicx,xcolor,physics,hyperref,tablefootnote,calrsfs}

\biboptions{comma,square,sort&compress}
\hypersetup{colorlinks = true, linkbordercolor = white, linkcolor = blue, citecolor = blue}

\journal{arXiv}

\newcommand{\rie}[1]{{\cal D}_{t}^{#1}}

\newcommand{\df}{D_{\rm f}\hspace{0.3mm}}
\newcommand{\prob}{{\cal P}}
\newcommand{\im}{{\rm i}}
\newcommand{\equa}[1]{Eq.~(\ref{#1})} \newcommand{\equas}[1]{Eqs.~(\ref{#1})}
\newcommand{\equass}[2]{Eqs.~(\ref{#1})-(\ref{#2})}
\newcommand{\equasa}[2]{Eqs.~(\ref{#1}) and (\ref{#2})}
\newcommand{\eqn}[2]{
\begin{align}
#1
\label{#2}
\end{align}
}
\newcommand{\ave}[1]{\hspace{-1pt}\left\langle{#1}\right\rangle\hspace{-1pt}}
\newcommand{\aveb}[1]{\bigg\langle{#1}\bigg\rangle}

\begin{document}

\begin{frontmatter}



\title{{\bf Rayleigh--B\'enard instability in a horizontal porous layer\\ with anomalous diffusion}}


\author{{\bf Antonio Barletta}}

\address{Department of Industrial Engineering. Alma Mater Studiorum Universit\`a di Bologna.\\ Viale Risorgimento 2, 40136 Bologna, Italy\\[3pt] \textrm{\tt antonio.barletta@{}unibo.it}\\
{\sf ORCID: 0000-0002-6994-5585}}

\begin{abstract}
The analysis of the Rayleigh-B\'enard instability due to the mass diffusion in a fluid-saturated horizontal porous layer is reconsidered. The standard diffusion theory based on the variance of the molecular position growing linearly in time is generalised to anomalous diffusion, where the variance is modelled as a power-law function of time. A model of anomalous diffusion based on a time-dependent mass diffusion coefficient is adopted, together with Darcy's law, for momentum transfer, and the Boussinesq approximation, for the description of the buoyant flow. A linear stability analysis is carried out for a basic state where the solute has a potentially unstable concentration distribution varying linearly in the vertical direction and the fluid is at rest. It is shown that any, even slight, departure from the standard diffusion process has a dramatic effect on the onset conditions of the instability. This circumstance reveals a strong sensitivity to the anomalous diffusion index. It is shown that subdiffusion yields instability for every positive mass diffusion Rayleigh number, while superdiffusion brings stabilisation no matter how large is the Rayleigh number. A discussion of the linear stability analysis based on the Galilei-variant fractional-derivative model of subdiffusion is eventually carried out. 
\end{abstract}


\end{frontmatter}


\section{Introduction}

Anomalous diffusion defines a class of processes where the statistics of molecular dynamics departs significantly from the standard Brownian motion theory, giving rise to such phenomena as superdiffusion or subdiffusion \cite{metzler2000random, DOSSANTOS201986}. Anomalous diffusion is often termed fractional diffusion when the governing equation involves fractional derivatives. Unlike classical diffusion, which is characterised by second-order spatial derivatives in space and first-order in time, fractional diffusion may be modelled by derivatives of non-integer order \cite{oldham1974fractional, hermannfractional}. In fact, the fractional derivative is a mathematical tool aimed to capture the memory effects and the long-range correlations typical of anomalous diffusion. Such effects may be present in a wide variety of physical, biological, and mathematical systems \cite{kulish2002application}. In fact, anomalous or fractional diffusion has been utilised to describe various geological processes, such as the spreading of contaminants in groundwater, the diffusion of gases in the atmosphere, and the heat and mass transfer in porous rocks. In particular, the transport of particles in complex media, such as porous materials, disordered systems, or fractal structures are typical situations where anomalous diffusion features might emerge \cite{sierociuk2013modelling}. 
When modelling a process of anomalous diffusion, the simplest approach does not necessarily involve fractional derivatives provided that the coefficient of diffusion, or diffusivity, is considered as time dependent \cite{wu2008propagators, henry2010introduction, sokolov2012models}.

At the molecular scale, the classical physical model of a diffusion process follows the random walk statistics where the variance in the random position of the molecules, $\sigma_x^2$, increases linearly with time, $t$, namely
\eqn{
\sigma_x^2 \sim D t ,
}{1}
where $D$ is a diffusion coefficient having necessarily the SI units m$^2$s$^{-1}$. 
The proportionality between $\sigma_x^2$ and $t$ is the ultimate reason underlying the well-known structure of the standard diffusion equation \cite{metzler2000random} with a first-order derivative in time and second-order derivatives in the coordinates leading to the Laplacian term, when the diffusion process is isotropic. Anomalous diffusion may be caused by a departure from the standard statistics of molecular random walk devised in the theory of Brownian motion \cite{henry2010introduction}. 
The characteristic behaviour of molecular random walk in anomalous diffusion is one where \equa{1} is generalised to
\eqn{
\sigma_x^2 \sim \df  t^r ,
}{2}
where $r$ is a positive dimensionless parameter which can be either in the range $r < 1$
(subdiffusion), or in the range $r > 1$ (superdiffusion), while $r=1$ brings back to \equa{1}, {\em i.e.}, to a standard diffusion process. In \equa{2}, $\df$ generalises $D$ as a fractional diffusion coefficient, where its SI units are m$^2$s$^{-r}$. An example of superdiffusion is when the anomalous behaviour
is caused by occasional events, or long molecular jumps, known as Levy flights \cite{dubkov2008levy} altering the local character of the diffusion process. In \equa{2}, one can think to a model where a time-dependent diffusion coefficient $D = \df r t^{r-1}$ is employed \cite{wu2008propagators, henry2010introduction, sokolov2012models}.

The Rayleigh-B\'enard instability in a horizontal fluid layer or fluid-saturated porous layer is the phenomenon of convective cellular flow that occurs in a system with a layer of fluid heated from below and cooled from above. Such a cellular flow pattern is possible only when the fluid layer is subjected to a sufficiently large temperature difference between the boundary planes or, in dimensionless terms, when the Rayleigh number is large enough \cite{straughan2013energy}. The mass diffusion version of the Rayleigh-B\'enard instability is also a possible phenomenon, where the mass diffusion of a solute substance replaces the heat diffusion as the activation mechanism for the convective cellular flow \cite{straughan2020heated}. With either heat diffusion or mass diffusion, the saturated porous medium version of the Rayleigh-B\'enard instability is also termed Horton-Rogers-Lapwood (HRL) problem \cite{horton1945convection, lapwood1948convection, nield1968, straughan2008stability, straughan2013energy, 
NiBe17,barletta2019routes}. In particular, the termohaline convection \cite{nield1968} identifies a situation where both heat and mass diffusion are present in determining the onset conditions for the cellular flow in the horizontal fluid-saturated porous medium. A recent paper by \citet{karani2017onset} explores the HRL problem by employing a model of heat diffusion where the advection term is expressed through a fractional gradient of the temperature field. Another recent study by \citet{klimenko2017effect} is focussed on the mass diffusion HRL problem with the diffusion equation expressed according to a two-phase fractional mobilisation/immobilisation model, where the immobilisation of the diffusing solute molecules means the adsorption of such molecules by the porous solid matrix. The more general field of microfluidic effects possibly under conditions of local thermal non-equilibrium has been widely investigated in the last decades \cite{banu2002onset, REES2005147, straughan2006global, rees2008local, barletta2012local, straughan2013porous,  straughan2015micro}. {Further recent studies deal with other relevant aspects of the mass diffusion interplay with convective flows \cite{10.1063/5.0056509, 10.1063/5.0059313, 10.1063/5.0069853, 10.1063/5.0139711, 10.1063/5.0153062}.}

The aim of this paper is to examine the effect of a departure from the standard mass diffusion theory, as modelled by an index $r \ne 1$ in \equa{2}, on the onset conditions for the solutal Rayleigh-B\'enard instability in a fluid-saturated porous layer. 
In Section~\ref{anodif}, very simple and intuitive arguments leading to the anomalous diffusion model, where the mass diffusivity is expressed as a power-law function of time, are proposed. Such a survey, not aimed at mathematical rigour or generality, is just meant as a quick tool for possible readers not specifically familiar with the topic. Then, in Section~\ref{mathform}, the mass diffusion equation is formulated by using a time-dependent diffusivity, while Darcy's law and the Boussinesq approximation are employed as a model of buoyant flow in the porous medium. It is well-known that, according to the standard mass diffusion theory, the onset conditions of the solutal instability in the layer are achieved when the mass diffusion Rayleigh number, $Ra$, lies above the minimum of the neutral stability curve, {\em i.e.}, for $Ra > 4\pi^2$ \cite{horton1945convection, lapwood1948convection, nield1968, straughan2008stability, NiBe17,barletta2019routes}. 
The basic stationary state with a zero velocity field is described in Section~\ref{bastst}. Then,
in Section~\ref{listan}, it is shown that a departure from the standard diffusion process, \equa{1}, leading to an anomalous process described by \equa{2} with $r \ne 1$, means a dramatic change in the onset conditions for the instability of the basic stationary state. In particular, $r < 1$ (subdiffusion) entails linear instability for every $Ra > 0$, while $r > 1$ (superdiffusion) means linear stability for every $Ra > 0$. This result implies an extreme sensitivity of the model to any departure from the standard statistics of the mass diffusion process. Finally, in Section~\ref{alfrdemo}, a discussion regarding the linear stability analysis and the onset conditions for the instability evaluated according to the Galilei-variant fractional-derivative model of subdiffusion \cite{metzler2000random} is presented.

\section{A na\"ive description of anomalous diffusion}\label{anodif}
In this section, we provide an informal description of the physics behind the phenomenon of anomalous diffusion. There is no aim at mathematical rigour or generality in the arguments reported, but only the objective of providing a short survey for the benefit of the reader not specifically familiar with the topic.
The forthcoming procedure is mostly a simplification of a similar layout proposed by many authors, such as \citet{henry2010introduction, sokolov2012models} and  \citet{POMEAU2017570}.

\subsection{The Langevin equation}

\begin{figure}[t]
\centering
\includegraphics[width=0.3\textwidth]{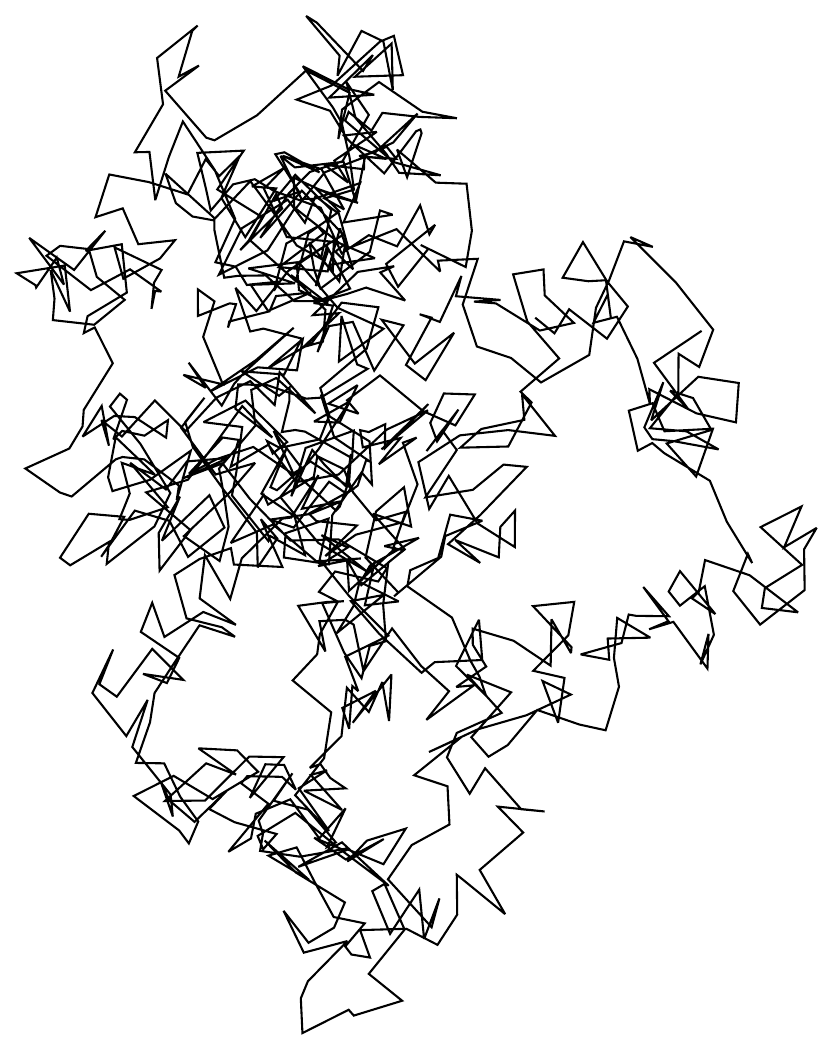}
\caption{\label{figrw}A qualitative sketch of a two-dimensional random walk.}
\end{figure}

The motion of a solute molecule into a milieu of base fluid molecules subject to thermal agitation can be described through the Langevin equation \cite{POMEAU2017570},
\eqn{
m  \dv[2]{x_i(t)}{t} = - \gamma  \dv{x_i(t)}{t} + F_i(t) ,
}{na1}
where $m$ is the mass of the molecule, $\gamma$ is a positive constant expressing the frictional coefficient and $F_i(t)$ is the $i$th component of the stochastic force acting on the molecule. The friction force $- \gamma  \dd x_i(t)/\dd t$ describes the collective dissipative effect of the molecular milieu, while the random force $F_i(t)$ expresses a stochastic signal accounting for the random character of the motion undergone by the solute molecule (see Fig.~\ref{figrw}). 

For the sake of simplicity and without any detriment of the physical argument, we will consider a one-dimensional motion meaning that the subscript $i$ in \equa{na1} will be hereafter omitted. 

The Langevin equation is the extension of Newton's second law of dynamics to a case where the motion is influenced by a random force, $F(t)$. The statistical properties of $F(t)$ are two: the average value $\ave{ F(t) }$ evaluated at any time $t$ over an ensemble of identically prepared solute molecules is zero; the stochastic function $F(t)$ is not correlated with the instantaneous position of the solute molecule, $x(t)$. In formulas, we have
\eqn{
\ave{ F(t) } = 0 \qc \ave{ x(t) F(t) } = \ave{ x(t) } \ave{ F(t) } = 0.
}{na2}
We evaluate the ensemble average $\ave{ \cdot }$ of the one-dimensional version of \equa{na1}, by using \equa{na2},
\eqn{
m \dv[2]{}{t} \ave{ x(t) } = - \gamma \dv{}{t} \ave{ x(t) }.
}{na2a}
Equation~({\ref{na2a}) yields
\eqn{
\dv{}{t} \ave{ x(t) } = \left. \dv{}{t} \ave{ x(t) } \right|_{t=0} e^{- \gamma t/m} .
}{na2b}
Hence, after a transient exponential decay, when $t \gg m/\gamma$, $\ave{ x(t) }$ can be approximated as a constant which can be set to zero by adjusting the origin of the $x$ axis,
\eqn{
\ave{x(t)} \approx 0   .
}{na2c}
Let us now multiply the one-dimensional version of \equa{na1} by $x(t)$ and evaluate its ensemble average. As a consequence of \equa{na2}, we obtain
\eqn{
m \ave{ x(t) \dv[2]{x(t)}{t} } = - \gamma \aveb{ x(t) \dv{x(t)}{t} },
}{na3}
Thus, by some straightforward mathematical manipulations, we can rewrite \equa{na3} as
\eqn{
m \dv{}{t}  \ave{  x(t) \dv{ x(t) }{t} } - m \aveb{ \qty[ \dv{ x(t)}{t} ]^2 } = - \gamma \aveb{ x(t) \dv{ x(t) }{t} }.
}{na4}
We can introduce the function 
\eqn{
Y(t) = \aveb{ x(t) \dv{ x(t) }{t} } = \frac{1}{2} \dv{}{t} \langle \qty[ x(t) ]^2 \rangle ,
}{na5}
so that \equa{na4} reads
\eqn{
m\, \dv{ Y(t) }{t} + \gamma Y(t) = m \aveb{ \qty[ \dv{ x(t)}{t} ]^2 } .
}{na6}
Standard diffusion theory is based on the assumption of thermodynamic equilibrium at a constant temperature $T$. Hence, following the equipartition theorem of statistical mechanics, one can express the average kinetic energy of the molecule, considered as a pointlike particle, in the form
\eqn{
\frac{m}{2} \aveb{ \qty[ \dv{ x(t)}{t} ]^2 } = \frac{1}{2} \kappa_B T,
}{na7}
where $\kappa_B = 1.380649 \times 10^{-23} \;{\rm J K^{-1}}$ is Boltzmann's constant. This means that the right hand side of \equa{na6} is a constant, so that this equation can be easily integrated,
\eqn{
Y(t) = \frac{\kappa_B T}{\gamma} + \qty[ Y(0) - \frac{\kappa_B T}{\gamma} ] e^{- \gamma t/m} .
}{na8}
Once the damping effect of the exponential in time is completed, {\em i.e.} when $t \gg m/\gamma$, the function $Y(t)$ turns out to be a constant. Thus, \equasa{na5}{na8} allow one to obtain
\eqn{
\langle \qty[ x(t) ]^2 \rangle - \langle \qty[ x(0) ]^2 \rangle\approx \frac{2 \kappa_B T}{\gamma} t.
}{na9}
We note that, from \equasa{na2c}{na9}, the variance $\sigma_x^2$ of the position $x(t)$ of the molecule at a given time $t$ can be expressed as
\eqn{
\sigma_x^2 = \langle \qty[ x(t) ]^2 \rangle - \ave{ x(t) }^2 = \langle \qty[ x(t) ]^2 \rangle \approx \frac{2 \kappa_B T}{\gamma} t ,
}{na10}
where it is implicitly assumed that the initial position of the solute molecule is known with certainty, {\em i.e.} $\langle \qty[ x(0) ]^2 \rangle = 0$.
One may introduce the diffusion coefficient $D$ defined as 
\eqn{
D = \frac{1}{2} \dv{\sigma_x^2}{t} \approx \frac{\kappa_B T}{\gamma} ,
}{na11}
so that \equa{na10} now reads
\eqn{
\sigma_x^2 = 2 D t .
}{na12}
This description accounts for standard diffusion processes. The anomalous diffusion deviates from this result due to the breakdown of the thermodynamic equilibrium assumption. The lack of thermodynamic equilibrium has several implications, such as the activation of memory effects and relaxation phenomena. Overall, the consequence is that the equipartition theorem cannot be invoked and the right hand side of \equa{na7} is not a constant any more. Thus, \equa{na7} now reads
%
\eqn{
\frac{m}{2} \aveb{ \qty[ \dv{ x(t)}{t} ]^2 } = \frac{1}{2} h(t),
}{na13}
where $h(t)$ is a positive-definite memory function. In other words, the instantaneous value of the average kinetic energy is not merely a function of the thermodynamic state, but a function of the stochastic process undergone so far by the solute molecule. Hence, \equasa{na6}{na7} are to be reformulated as
\eqn{
m\, \dv{ Y(t) }{t} + \gamma Y(t) = h(t) .
}{na14}
%
The simplest model with a non-constant memory function $h(t)$ is one where \equa{na14} entails $Y(t)$ 
to be given by a power-law function of $t$ at sufficiently large times $(t \gg m/\gamma)$,
\eqn{
Y(t) = \df r t^{r-1} ,
}{na15b}
where $\df$ and $r$ are real positive constants. Relying on \equasa{na5}{na15b}, one has
\eqn{
\langle \qty[ x(t) ]^2 \rangle - \langle \qty[ x(0) ]^2 \rangle= 2 \df t^{r},
}{na16}
instead of \equa{na9} and 
\eqn{
\sigma_x^2 = 2 \df t^{r} .
}{na17}
instead of \equa{na12}. Thus, one can make \equasa{na12}{na17} formally equivalent by introducing a time-dependent diffusion coefficient defined as
\eqn{
D = \frac{1}{2} \dv{\sigma_x^2}{t} = \df r t^{r-1} .
}{na18}
On average, the small step-sizes in position and time $\Delta x$ and $\Delta t$ during the molecular random walk can be approximated by employing the derivative $\dd \sigma_x^2/ \dd t$, namely
\eqn{
\Delta x^2 \approx \dv{\sigma_x^2}{t} \Delta t = 2 D \Delta t.
}{na19}

\begin{figure}[t]
\centering
\includegraphics[width=0.7\textwidth]{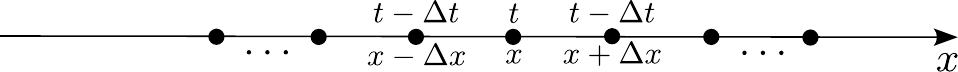}
\caption{\label{figme}Illustration of the steps undergone by the one-dimensional random walker in the time interval $[t-\Delta t, t]$.}
\end{figure}

\subsection{The master equation}
Let us consider the probability density function $\prob(x,t)$. The probability for the randomly walking molecule to be localised in an infinitesimal range between $x$ and $x+\dd x$ at time $t$ is given by $\prob(x,t) \dd x$. Then, by employing the average step-sizes defined by \equa{na19}, one can write
\eqn{
\prob(x,t) \approx \frac{1}{2} \qty[ \prob(x - \Delta x, t - \Delta t) + \prob(x + \Delta x, t - \Delta t) ] ,
}{na20}
where the approximation improves as the step-sizes $\Delta x$ and $\Delta t$ become smaller and smaller. One can figure out \equa{na20} by realising that the molecule is at $(x,t)$ if,  at the preceding instant $t - \Delta t$, its position is one step backward or one step forward from $x$ (see Fig.~\ref{figme}). Here, the two possibilities are statistically equivalent and, hence, weighted with $1/2$. There is a more rigorous version of \equa{na20} where $\Delta x$ and $\Delta t$ are considered as statistical variables with their own probability density \cite{metzler2000random, henry2010introduction}, but we will not discuss this approach here as it does not sensibly affect the conclusions to be achieved. We now employ a Taylor series expansion at the lowest orders to write
\eqn{
\prob(x \pm \Delta x, t - \Delta t) \approx \prob(x, t) \pm \pdv{\prob(x,t)}{x} \Delta x 
-  \pdv{\prob(x,t)}{t} \Delta t + \frac{1}{2}  \pdv[2]{\prob(x,t)}{x} \Delta x^2
\nonumber\\
 \mp \pdv[2]{\prob(x,t)}{x}{t} \Delta x\, \Delta t + \frac{1}{2}  \pdv[2]{\prob(x,t)}{t} \Delta t^2 
 \pm \order{\Delta x^3}  + \order{\Delta x^2\Delta t}  \pm  \order{\Delta x\,\Delta t^2} + \order{\Delta t^3} .
}{na21}
By substituting \equa{na21} into \equa{na20}, we obtain
\eqn{ 
\pdv{\prob(x,t)}{t} \Delta t \approx 
 \frac{1}{2}  \pdv[2]{\prob(x,t)}{x} \Delta x^2 
 + \frac{1}{2}  \pdv[2]{\prob(x,t)}{t} \Delta t^2 + \order{\Delta x^2\Delta t} + \order{\Delta t^3} .
}{na22}
The approximation becomes equality when $\Delta x$ and $\Delta t$ tend to zero, so that \equa{na19} provides a very accurate evaluation of the ratio $\Delta x^2/\Delta t$. We divide \equa{na22} by $\Delta t$ and we take the limits $\Delta t \to 0$ and  $\Delta x \to 0$. Thus, by using \equa{na19}, we obtain the master equation \cite{sokolov2012models}
\eqn{ 
\pdv{\prob}{t} =D \pdv[2]{\prob}{x},
}{na23}
with $D$ expressed by \equa{na18}.
The product between $\prob(x,t)$ and the overall mass ${\cal M}$ of the solute substance undergoing the random walk in the solvent molecular milieu yields the concentration field $C(x,t)$ of the solute. Thus, the master equation (\ref{na23}) is rewritten as the one-dimensional diffusion equation
\eqn{ 
\pdv{C}{t} =D \pdv[2]{C}{x}.
}{na24}
Having in mind the physics behind this simple one-dimensional argument, the generalisation to three dimensions and to the case where the underlying molecular milieu is subject to a deterministic velocity field with $j$th component $u_j$ is straightforward, so that \equa{na24} is written as
\eqn{ 
\pdv{C}{t} + u_j \pdv{C}{x_j} = D \laplacian{C}.
}{na25}
In \equa{na25} and thereafter, Einstein's notation for the sum over a repeated index is employed. Going three-dimensional also means that the average kinetic energy in \equa{na7} must include the contribution of the three square derivatives of $x(t)$, $y(t)$ and $z(t)$. Hence, as a consequence of the equipartition theorem relative to a pointlike particle, the constant $\kappa_B T/2$ on the right hand side of \equa{na7} must be multiplied by the number of degrees of freedom, namely it is to be replaced by $3\kappa_B T/2$. 

\section{Mathematical formulation}\label{mathform}
Let us consider a horizontal fluid-saturated porous layer with thickness $H$. The $z$ axis is chosen as vertical, so that the gravitational acceleration is $\vb{g} = - g \vu{e}_z$ with $\vu{e}_z$ the unit vector of the $z$ axis and $g$ the modulus of $\vb{g}$. As sketched in Fig.~\ref{figpm}, the origin of the $z$ axis is at the lower boundary plane, $z=0$, modelled as both impermeable and with a uniform solute concentration $C_1$. The upper boundary plane, $z=H$, is also impermeable, but kept at a uniform concentration $C_2 \ne C_1$. The width of the layer along the horizontal axes $x$ and $y$ is considered as infinite.

\subsection{Anomalous mass diffusion in a fluid-saturated porous medium}
The standard local mass diffusion equation of a solute in a fluid-saturated porous medium is formulated, starting from \equa{na25}, as \cite{NiBe17,barletta2019routes}
\eqn{
\phi \pdv{C}{t} + u_j \pdv{C}{x_j} - \phi D \nabla^2 C = 0 ,
}{3}
with a constant $D$.
We emphasise that $u_j$ now denotes the $j$th component (with $j = 1,2,3$, or $x,y,z$) of the seepage, or Darcy's, velocity, while $\phi$ is the porosity of the solid medium and $x_j$ stands for the $j$th Cartesian coordinate. The generalisation to anomalous diffusion of \equa{3} is obtained by using \equa{na18}, namely
\eqn{
\phi \pdv{C}{t} + u_j \pdv{C}{x_j} - \phi \df r t^{r-1} \nabla^2 C = 0 ,
}{4}
identifying the standard diffusion as the special case $r=1$ with $\df = D$. 

\begin{figure}[t]
\centering
\includegraphics[width=0.8\textwidth]{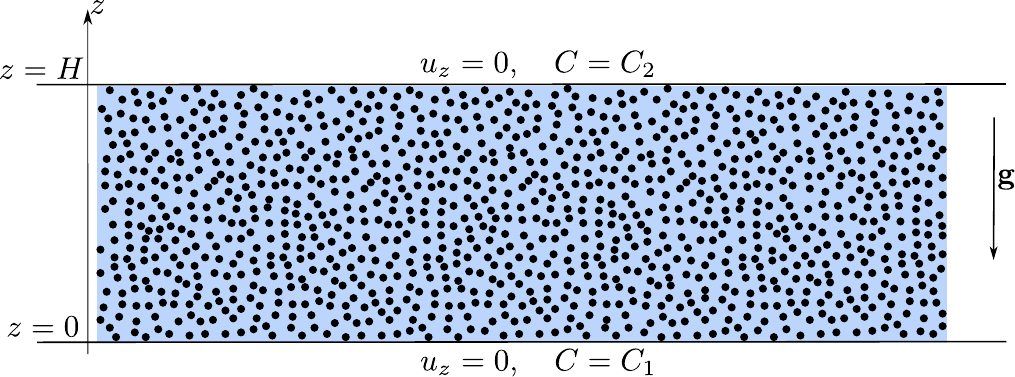}
\caption{\label{figpm}A sketch of the fluid-saturated porous layer and of the boundary conditions.}
\end{figure}

\subsection{Mass balance and momentum balance}
The model of fluid flow in a porous medium is based on Darcy's law \cite{NiBe17,barletta2019routes}, where the buoyancy force due to the solutal diffusion is taken into account by adopting the Boussinesq approximation \cite{NiBe17,barletta2019routes,barletta2022boussinesq,barletta2023use}. The local mass and momentum balance equations are thus given by
\eqn{
\pdv{u_j}{x_j} &= 0, \label{5}\\
\frac{\mu}{K} u_i &= - \pdv{p}{x_i} + \rho g \alpha \qty(C - C_0) \delta_{3i} , 
}{6}
where $p$ is the local difference between the pressure and the hydrostatic pressure, $\mu$, $\rho$ and $\alpha$ are the dynamic viscosity, the density and the solutal expansion coefficient of the fluid, respectively, while $K$ is the permeability of the porous medium and $\delta_{3i}$ is the $3i$th component of Kronecker's delta. The reference concentration, $C_0$, employed for the definition of the buoyancy force in the local momentum balance \equa{6} is chosen as that prescribed at the top boundary, namely $C_0 = C_2$. 

\subsection{Boundary conditions}
The boundary conditions are meant to describe impermeability and uniform solute concentrations at $z=0,H$. They are expressed as
\eqn{
z = 0: \qquad  &u_z = 0 \qc C = C_1,  \nonumber \\
z = H: \qquad  &u_z = 0 \qc C = C_2 .
}{7}

\renewcommand{\tabcolsep}{0.8cm}

\begin{table}[t]
\renewcommand{\tabcolsep}{0.8cm}
\centering 
\caption{Scales adopted for the dimensionless analysis}\vspace{1ex} 
\begin{tabular}{l c } 
\hline\hline 
Quantity & Scale \\  
\hline 
\vspace{-2ex}\\
coordinate & $H$ \\ 
time & $H^{2/r} \df^{-1/r}$ \\  
velocity & $H^{(r-2)/r} \df^{1/r}$ \\ 
pressure & $\mu K^{-1} H^{2(r-1)/r} \df^{1/r}$ \\ 
concentration$^*$ & $C_1 - C_2$ \\ 
[0.5ex] 
\hline 
\end{tabular}\\
$^*${\scriptsize The dimensionless concentration is defined as $\qty(C - C_2)/\qty(C_1 - C_2)$}
\label{tab1} 
\end{table}

\subsection{Dimensionless quantities}
The governing equations and boundary conditions (\ref{4}), (\ref{5})-(\ref{7}) can be expressed in a dimensionless form by adopting the scales defined in Table~\ref{tab1} and by employing, for the sake of simplicity, the same symbols for the dimensionless and dimensional quantities,
\eqn{
&\pdv{u_j}{x_j} = 0, \label{8}\\
&u_i + \pdv{p}{x_i} - \phi Ra C \delta_{3i} = 0 , \label{9}\\
&\phi \pdv{C}{t} + u_j \pdv{C}{x_j} - \phi r t^{r-1} \nabla^2 C = 0 , \label{10}\\
&z = 0: \qquad  u_z = 0 \qc C =1 ,  \nonumber \\
&z = 1: \qquad  u_z = 0 \qc C = 0 ,
}{11}
where $Ra$ in \equa{9} denotes the mass diffusion Rayleigh number defined as
\eqn{
Ra = \frac{\rho g \alpha \qty(C_1 - C_2) K H^{(2-r)/r}}{\phi \mu \df^{1/r}} .
}{12}
%
Hereafter, the Rayleigh number will be considered as positive or, equivalently, $\alpha \qty(C_1 - C_2)$ will be considered as positive, so that a downward dimensionless concentration gradient identifies a potentially unstable solutal stratification.

\section{The basic stationary state}\label{bastst}
As in the classical analysis of the solutal Rayleigh-B\'enard problem in a saturated porous medium, the basic stationary state is a zero velocity solution of \equass{8}{11} with a distribution of $C$ depending just on the $z$ coordinate,
\eqn{
u_i = 0 \qc C = 1 - z  \qc p = \phi Ra \qty(1 - \frac{z}{2}) z ,
}{13}
where the value of $p$ at $z=0$ has been conventionally set to zero.
Equation~(\ref{13}) describes a stationary motionless state with a potentially unstable vertical stratification of the concentration field $C$.

\section{The linear stability analysis}\label{listan}
Let us now investigate the dynamics for the small-amplitude perturbations of the basic state, \equa{13}, by defining
\eqn{
u_i = \epsilon U_i  \qc  C = 1 - z + \epsilon \chi \qc p = \phi Ra \qty(1 - \frac{z}{2}) z + \epsilon P ,
}{14}
where $\epsilon \ll 1$ is a positive perturbation parameter and $\qty(U_i, \chi, P)$ are the perturbation functions. Let us substitute \equa{14} into \equass{8}{11} by neglecting terms $\order{\epsilon^2}$, so that we obtain
\eqn{
&\pdv{U_j}{x_j} = 0, \label{15}\\
&U_i + \pdv{P}{x_i} - \phi Ra \chi \delta_{3i} = 0 , \label{16}\\
&\phi \pdv{\chi}{t} - U_z - \phi r t^{r-1} \nabla^2 \chi = 0 , \label{17}\\
&z = 0,1: \qquad  U_z = 0 \qc \chi = 0  .
}{18}
The eigenvalue problem (\ref{15})-(\ref{18}) is invariant under rotations around the vertical $z$ axis, so that we can focus on the $y$-independent solutions. By this argument, the two-dimensional analysis of \equass{15}{18} yields
\eqn{
&\pdv{U_x}{x} + \pdv{U_z}{z} = 0, \label{19}\\
&\pdv{U_x}{z} - \pdv{U_z}{x} + \phi Ra \pdv{\chi}{x} = 0 , \label{20}\\
&\phi \pdv{\chi}{t} - U_z - \phi r t^{r-1} \nabla^2 \chi = 0 , \label{21}\\
&z = 0,1: \qquad  U_z = 0 \qc \chi = 0  ,
}{22}
where the $x$ and $z$ components of \equa{16} have been rearranged in a vorticity formulation to encompass the dependence on $P$, while the $y$ component of \equa{16} just yields $U_y=0$. Let us now introduce a streamfunction, $\psi$, defined as
\eqn{
U_x = \pdv{\psi}{z} \qc U_z = - \pdv{\psi}{x},
}{23}
so that \equa{19} is identically satisfied and \equass{20}{22} yield
\eqn{
&\laplacian{\psi} + \phi Ra \pdv{\chi}{x} = 0 , \label{24}\\
&\phi \pdv{\chi}{t} + \pdv{\psi}{x} - \phi r t^{r-1}\nabla^2 \chi = 0 , \label{25}\\
&z = 0,1: \qquad  \pdv{\psi}{x} = 0 \qc \chi = 0  .
}{26}

\subsection{The normal modes}
We now express $\psi$ and $\chi$ as normal modes
\eqn{
\psi = A(t) e^{\im k x} \sin(n\pi z) \qc \chi = B(t) e^{\im k x} \sin(n\pi z), \qfor n = 1, 2, 3, \ldots ,
}{27}
and $k$ denoting the wavenumber. Substitution of \equa{27} into \equasa{24}{25} yields
\eqn{
&A(t) = \frac{\im k \phi Ra}{k^2 + n^2\pi^2} B(t), \label{28}\\
&\dv{B(t)}{t} - \qty[ \frac{k^2 Ra}{k^2 + n^2 \pi^2} - r t^{r-1} \qty(k^2 + n^2 \pi^2) ] B(t) = 0,
}{29}
while the boundary conditions (\ref{26}) are identically satisfied. Equation~(\ref{28}) and integration of \equa{29} lead to 
\eqn{
&
\qty{ 
\mqty{A(t)\\B(t)}
} = 
\qty{ 
\mqty{A(0)\\B(0)}
} \exp\!\qty[ \qty(k^2 + n^2 \pi^2) G(t) ], \label{30}\\
&G(t) = S t - t^r , \qfor S = \frac{k^2 Ra}{\qty(k^2 + n^2 \pi^2)^2}.
}{31}
Function $G(t)$ is the key to tell stability from instability, by inspecting its behaviour at large times. 

\subsection{Standard diffusion $(r=1)$}
It is evident from \equasa{30}{31} that, with $r=1$, parameter $S$ determines the monotonic increasing or decreasing character of $G(t)$ and, hence, the growth $(S > 1)$or decay $(S < 1)$ in time of the perturbation amplitudes, $A(t)$ and $B(t)$. The threshold condition to instability is $S=1$, yielding the well-known neutral stability condition \cite{horton1945convection, lapwood1948convection, nield1968, NiBe17,barletta2019routes},
\eqn{
Ra = \frac{\qty(k^2 + n^2 \pi^2)^2}{k^2} ,
}{32}
where the minimum value of $Ra$, {\em i.e.} the critical value $Ra_c = 4 \pi^2$, is achieved with $n=1$ and $k = k_c = \pi$. No instability is possible when $Ra$ is subcritical.

\begin{figure}[t]
\centering
\includegraphics[width=0.8\textwidth]{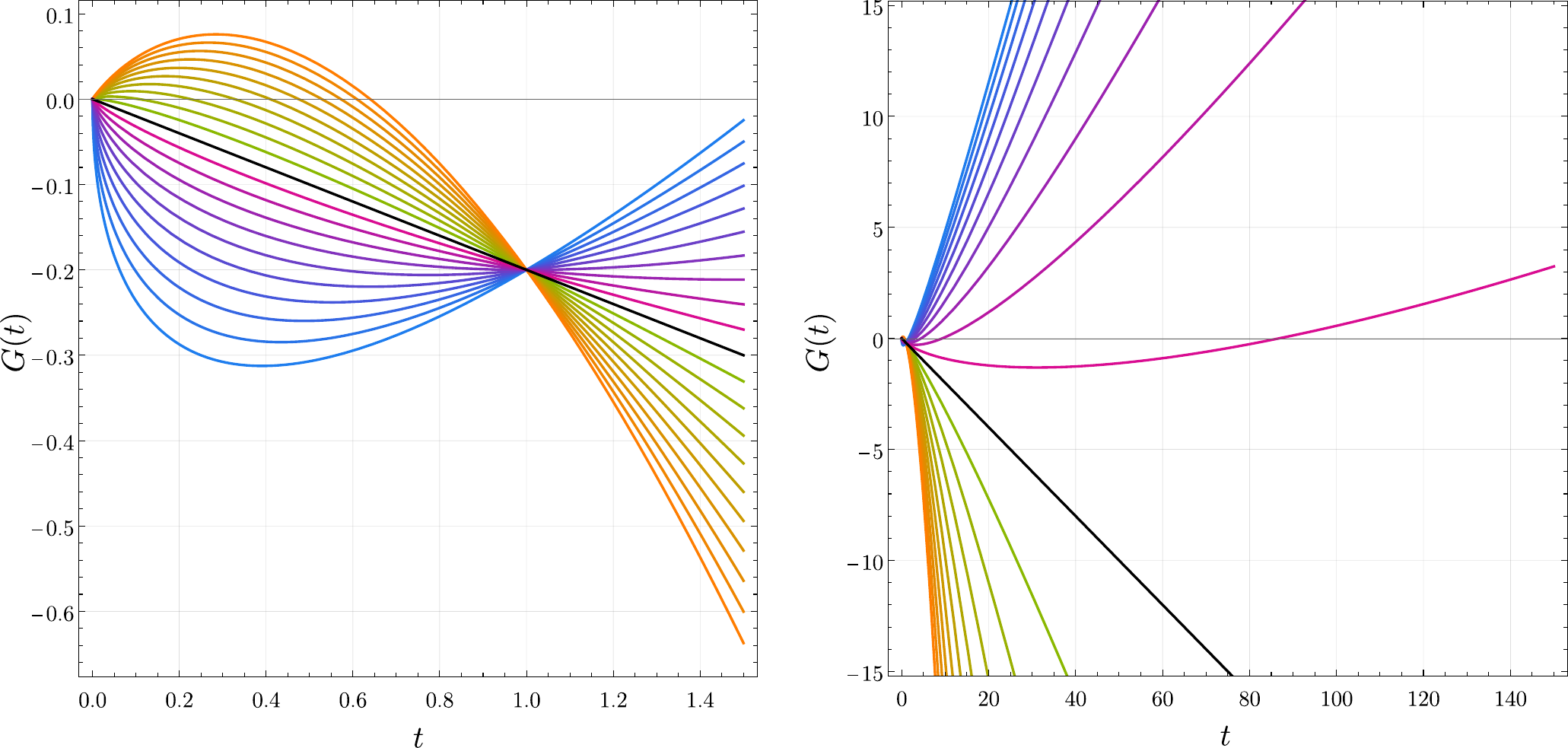}
\caption{\label{fig1}Plots of $G(t)$ for $S=0.8$. Colours from blue to magenta denote subdiffusion with values of $r$ from $0.5$ to $0.95$ in steps of $0.05$.
Colours from green to orange denote superdiffusion with values of $r$ from $1.05$ to $1.5$ in steps of $0.05$. The black line is for standard diffusion, $r=1$.}
\end{figure}

\begin{figure}[h!!]
\centering
\includegraphics[width=0.8\textwidth]{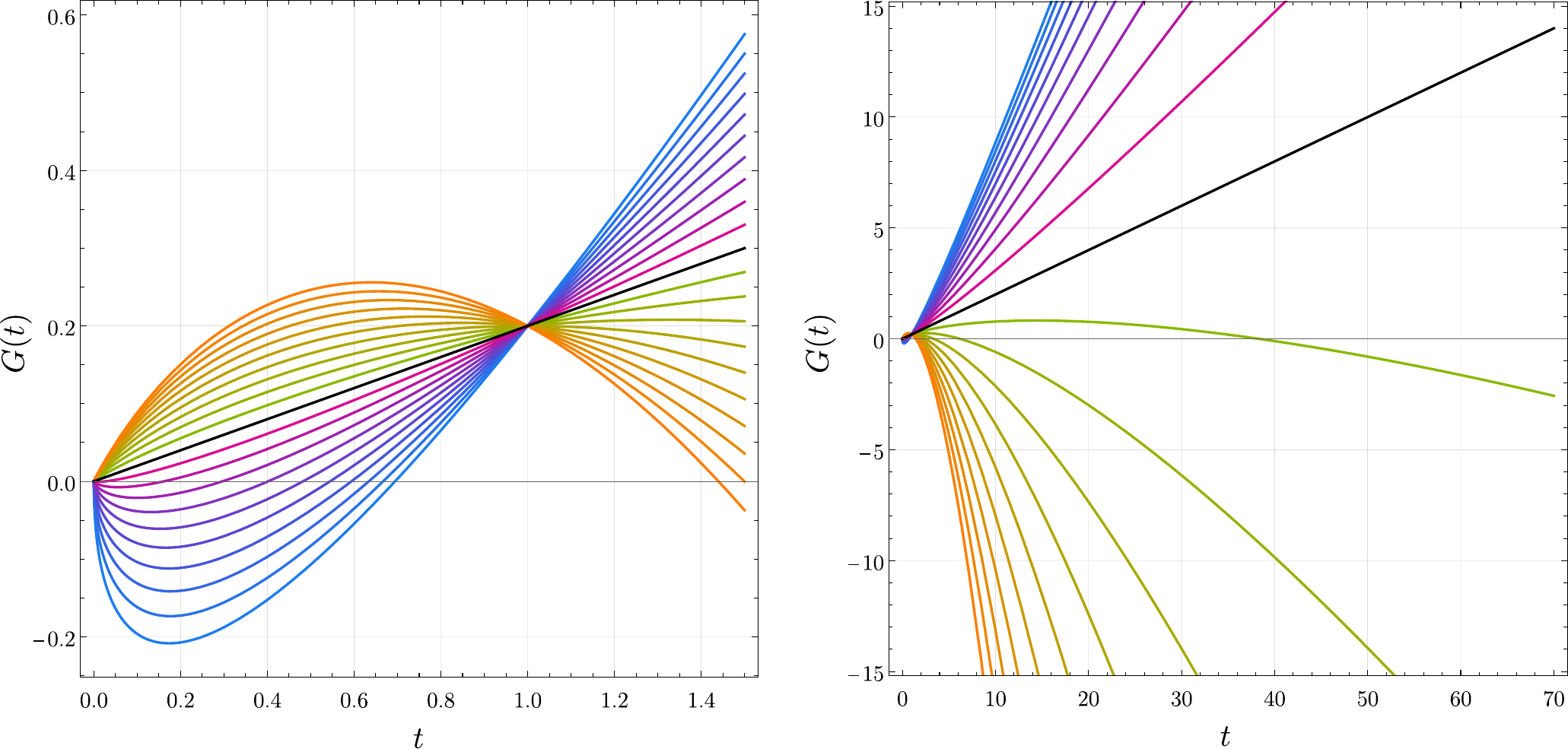}
\caption{\label{fig2}Plots of $G(t)$ for $S=1.2$. Colours from blue to magenta denote subdiffusion with values of $r$ from $0.5$ to $0.95$ in steps of $0.05$.
Colours from green to orange denote superdiffusion with values of $r$ from $1.05$ to $1.5$ in steps of $0.05$. The black line is for standard diffusion, $r=1$.}
\end{figure}

\subsection{Subdiffusion $(r<1)$ and superdiffusion $(r>1)$}\label{subsuper}
With $r \ne 1$, parameter $S$ has some importance only at the early stages of the evolution in time of $G(t)$. At large times, $G(t)$ is a monotonic increasing or decreasing function depending only on $r$ being smaller or larger than $1$, respectively. Hence, the perturbations, $A(t)$ and $B(t)$,  at large times undergo an unbounded growth for subdiffusion $(r < 1)$ or a decay to zero for superdiffusion $(r > 1)$,
\eqn{
\lim_{t \to +\infty} 
\qty{ 
\mqty{A(t)\\B(t)}
}
 = 
\begin{cases}
\infty &\qfor r < 1 \\
0 &\qfor r > 1
\end{cases} \qquad \forall S > 0 .
}{33}
Incidentally, whenever $S \le 0$, meaning $Ra \le 0$, \equasa{30}{31} lead to the prediction of a decay to zero of the perturbations independently of $r$. This result is quite reasonable given that $S \le 0$ addresses cases of a stably stratified fluid in the basic state.

Figures~\ref{fig1} and \ref{fig2} display the trend of function $G(t)$ for either $S=0.8$ (Fig.~\ref{fig1}) or $S = 1.2$ (Fig.~\ref{fig2}). Several values of $r$ are spanned in such figures describing both subdiffusion and superdiffusion, with a smaller or a larger time range starting from $t=0$ in the left hand frame and in the right hand frame, respectively. Figure~\ref{fig1} illustrates a case where standard diffusion predicts stability, $S < 1$. On the other hand, Fig.~\ref{fig2} is relative to a case where standard diffusion would yield instability, $S > 1$. 

There is a simple heuristic argument explaining why $r < 1$ yields instability under conditions where standard diffusion predicts stability, {\em i.e.}, for $0 < S < 1$. As pointed out by several authors, as for instance \citet{normand1977convective}, the Rayleigh-B\'enard instability results from the synergic action of three physical effects: the buoyancy, the diffusion and the viscous friction. The buoyancy force is powered by an inefficient diffusion inasmuch as such inefficiency is responsible for locally large concentration gradients. The initiating convection cells undergo an amplification in time, leading to instability, only if the buoyancy force is so intense as to prevail over the viscous damping. Subdiffusion $(r<1)$ makes the time-dependent diffusivity, $D = \df r t^{r-1}$, smaller and smaller as time increases, thus creating conditions where the buoyancy force is increasingly intense with time. Thus, a time exists where the buoyancy force ends up to prevail over the viscous damping and the instability starts. A similar argument can easily be used to justify why superdiffusion $(r>1)$ yields linear stability even when standard diffusion would predict instability $(S>1)$. For $r>1$, a time always exists where the diffusivity is so large and the buoyancy force becomes consequently so weak that the viscous damping eventually prevails, thus switching off the initiating convection cells.

With $r \ne 1$, \equa{31} shows that function $G(t)$ has not a monotonic behaviour for $t > 0$. In fact, its derivative vanishes at time $t = \qty(S/r)^{1/(r-1)}$ meaning a change from an increasing trend to a decreasing trend or vice versa. More precisely, $t = \qty(S/r)^{1/(r-1)}$ is a minimum for subdiffusion $(r<1)$ and a maximum for superdiffusion $(r>1)$. For subdiffusion, the transient damping at earlier times followed by an asymptotic unbounded growth at large times is a trend unlikely modified by the nonlinearity of the perturbation dynamics if not for a possible nonlinear saturation eventually bounding the linearly unbounded growth. Things might be different for superdiffusion where the transient growth at earlier times can mean a strong initial amplification of the initial perturbation amplitudes $A(0)$ and $B(0)$. This transient growth could drive the evolution from an initial linear behaviour to a nonlinear one before reaching the maximum of the amplitudes. The possible emergence of such a nonlinearity for $t \le \qty(S/r)^{1/(r-1)}$ might hinder the linear predictions for the damping of the amplitudes $A(t)$ and $B(t)$ at larger times.

\section{An alternative fractional derivative model}\label{alfrdemo}
An alternative model of subdiffusion $(r<1)$ could be employed based on the use of the fractional time derivative in the mass balance equation for the solute, so that \equa{10} is replaced by 
\eqn{
\phi \rie{r} C + \xi u_j \pdv{C}{x_j} - \phi \nabla^2 C = 0 ,
}{63}
where $\rie{r}$ is the fractional time derivative of order $r$.

Equation~(\ref{63}) is grounded on the Galilei-variant fractional diffusion-advection equation \cite{compte1998fractional, metzler2000random}, where $\xi$ is an advection-mobility dimensionless parameter, based on an advection-mobility time constant $\tau$, defined as
\eqn{
\xi = \qty[ \qty(\frac{\df}{H^2})^{1/r} \tau ]^{1-r}.
}{64}
The quantity $\tau$ is needed in the model for dimensional consistency reasons whenever $r \ne 1$.
Compatibility with standard diffusion is assured as, for $r \to 1$, \equa{64} yields $\xi \to 1$, whatever is the value of $\tau$. 
Furthermore, 
$\rie{r} f(t)$ is assumed to be the Caputo derivative of a function $f(t)$, defined as \cite{henry2010introduction}
\eqn{
\rie{r} f(t) = \frac{1}{\Gamma(1-r)}  \int_0^t \qty(t-s)^{-r}\, \dv{f(s)}{s}\, \dd s,  \qfor 0 < r < 1,
}{65}
{where $\Gamma\qty(\cdot)$ is Euler's Gamma function.}
With this approach, there is no change in the determination of the basic state as \equa{13} still yields a solution of \equas{8}, (\ref{9}), (\ref{11}) and (\ref{63}). This is a consequence of the property that the Caputo derivative of a time-independent quantity is zero, a property not satisfied by other definitions of the fractional time derivative \cite{hermannfractional}.

\begin{figure}[t]
\centering
\includegraphics[width=0.4\textwidth]{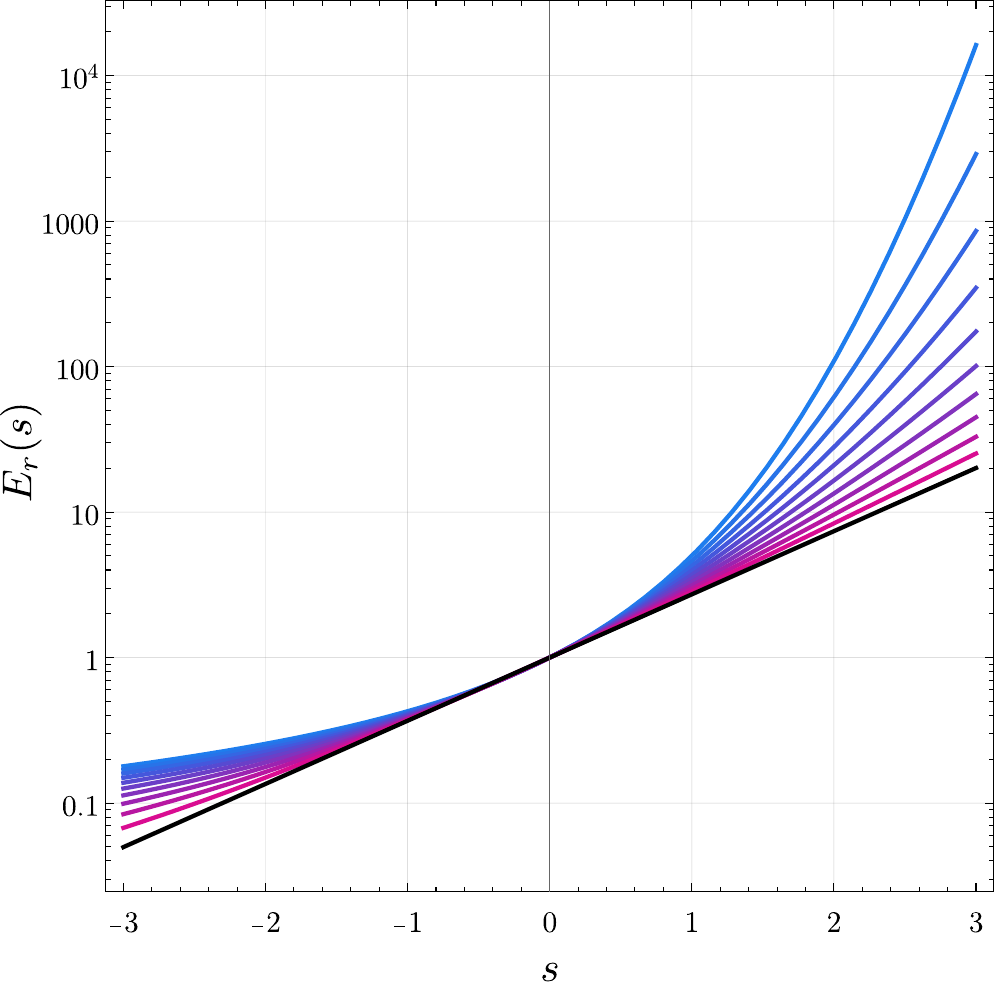}
\caption{\label{fig3}Plots of the Mittag-Leffler function $E_r(s)$ versus $s$. Coloured lines from blue to magenta are for values of $r$ from $0.5$ to $0.95$ in steps of $0.05$, while the black line is for $r=1$.}
\end{figure}

On perturbing the basic state given by \equa{13}, \equasa{24}{26} are left unaltered, while \equa{25} is now replaced by
\eqn{
&\phi \rie{r} \chi + \xi \pdv{\psi}{x} - \phi \nabla^2 \chi = 0 .
}{66}
The use of the normal modes given by \equa{27} leads to \equa{28} and
\eqn{
\rie{r}{B(t)} - \qty[ \frac{k^2 \xi Ra}{k^2 + n^2 \pi^2} - \qty(k^2 + n^2 \pi^2) ] B(t) = 0,
}{67}
instead of \equa{29}. The solution of \equasa{28}{67} can be expressed in terms of the Mittag-Leffler function $E_r\qty(\cdot)$ \cite{duan2013eigenvalue},
\eqn{
&
\qty{ 
\mqty{A(t)\\B(t)}
} = 
\qty{ 
\mqty{A(0)\\B(0)}
} E_r \qty( \lambda t^r) \qfor \lambda = \frac{k^2 \xi Ra}{k^2 + n^2 \pi^2} - \qty(k^2 + n^2 \pi^2) .
}{68}
The Mittag-Leffler function $E_r\qty(\cdot)$ is defined as \cite{duan2013eigenvalue},
\eqn{
E_r (s) = \sum_{n=0}^{\infty} \frac{s^n}{\Gamma(r n +1)} \qc \qfor s \in \mathbb{C} .
}{69}
It is evident from \equa{69} that the Mittag-Leffler function can be considered as a fractional extension of the exponential function. In fact, $E_1 (s) = e^s$. Just as the exponential function, $E_r (s)$ monotonically increases with $s \in \mathbb{R}$ \cite{gorenflo2019mittag} as illustrated in Fig.~\ref{fig3}. Function $E_r (s)$, for $1/2 \le r < 1$, is compared in this figure with $E_1 (s) = e^s$. For $1/2 \le r < 1$, function $E_r (s)$ grows significantly faster than the exponential with increasing $s$, for $s>0$. On the other hand, it decays to zero slower than the exponential with increasing $|s|$, for $s<0$. From these properties of the Mittag-Leffler function, one is led to the conclusion that instability happens for $\lambda > 0$. Then, by taking into account \equa{68}, the neutral stability condition $(\lambda = 0)$ occurs with the $n=1$ modes, provided that 
\eqn{
Ra = \frac{\qty(k^2 + \pi^2)^2}{\xi k^2} .
}{70}
The smaller is $\xi$ the lower are the neutral stability values of $Ra$. At the very least, the neutral stability condition arises when $k$ and $Ra$ coincide with the critical values
\eqn{
k_c = \pi \qc Ra_c = \frac{4 \pi^2}{\xi} .
}{71}
How such conditions for instability compare with those for standard diffusion $(r=1, \xi=1)$ depends entirely on the value of $\xi$, which is a parameter to be determined experimentally. A possible agreement between the predictions gathered from the time-dependent diffusivity model established in  Section~\ref{subsuper} and the predictions grounded on \equasa{70}{71} suggests that $Ra_c$ for $1/2 \le r < 1$ should be smaller than $4 \pi^2$, which is the critical value for the standard diffusion case $(r=1, \xi=1)$. In fact, in  Section~\ref{subsuper}, we proved that subdiffusion yields a destabilising effect with respect to standard diffusion.
This argument could be corroborated only if subdiffusion described through \equa{63} means $\xi > 1$. However, $\xi$ is based on the advection-mobility time parameter $\tau$ which, as suggested by \citet{PhysRevE.55.6821}, ``must be handled as a macroscopic parameter on an equal footing with'' the fractional diffusivity, $\df$. As such, only experimental data (currently unavailable to the best of the author's knowledge) may shed some light on the value of $\tau$ under given circumstances, and the experimental analysis of the Rayleigh-B\`enard problem in a porous material may be a case where an indirect measurement of $\tau$ can be accomplished through a measurement of $Ra_c$. 

If a separate determination of $(r, \df, \tau)$ serves to establish the effect of subdiffusion on the onset of the instability, when subdiffusion is modelled through \equa{63}, one can ground the instability predictions on a single parameter, $\xi Ra$. This product can be interpreted as a modified Rayleigh number $Ra_{\rm mod}$ which, from \equasa{12}{64}, is expressed as
\eqn{
Ra_{\rm mod} = \xi Ra = \frac{\rho g \alpha \qty(C_1 - C_2) K H}{\phi \mu \df \tau^{r-1}} .
}{72}
One can realise that $Ra_{\rm mod}$ is nothing but the Rayleigh number for standard diffusion in a porous medium when the diffusion coefficient $D$ is given by $\df \tau^{r-1}$. Interestingly enough, the predicted threshold condition to instability is all about the value of $\df \tau^{r-1}$ for subdiffusion as compared to the value of the mass diffusivity $D$ for standard diffusion. 
On purely physical grounds, this outcome can be compared with what has been argued in Section~\ref{subsuper}. We have gone into a constant diffusivity $D = \df \tau^{r-1}$ instead of a time-dependent diffusivity $D = \df r t^{r-1}$. The formal analogy turns out to have significant effects. In fact, the parameter $\tau$ does not change during the evolution in time of the perturbations. 
As a consequence, the diffusivity $D = \df r t^{r-1}$ decreases during the perturbation evolution, while the value of $D = \df \tau^{r-1}$ remains constant.

\section{Conclusions}
The onset of the Rayleigh-B\'enard instability in a horizontal porous layer driven by a solute concentration gradient has been studied by assuming an anomalous diffusion process. The underlying molecular dynamics for the solute species has been described in terms of a Brownian motion. The positions of the random walker at a given instant of time $t$ have a statistical distribution whose variance increases in time with a power law, $t^r$. The positive real index $r$ modulates the departure from standard diffusion processes, where the latter are characterised by $r=1$. Anomalous phenomena may be associated with either $r < 1$ (subdiffusion) or $r > 1$ (superdiffusion). 

The deduction of the balance equation for the anomalous diffusion based on a time-dependent diffusivity coefficient, $D = \df r t^{r-1}$ where $\df$ is a constant, has been surveyed and outlined. Then, the governing equations for the buoyant mass-diffusion flow in a fluid-saturated porous medium have been written. The basic stationary state with zero velocity and a uniform vertical gradient of the solute concentration has been defined, showing that such a basic state does not depend on the anomalous diffusion index $r$.

The linear stability analysis of the basic stationary state has been studied by expressing the perturbations in terms of normal modes with a time-dependent amplitude.  For standard mass diffusion, there exists a critical Rayleigh number, $Ra_c = 4\pi^2$, marking the transition to instability. On the other hand, the main features of the linear stability analysis for anomalous diffusion are the following:
\begin{enumerate}
\renewcommand{\labelenumi}{\theenumi)}
\item\label{it1} The evolution in time of the perturbation amplitudes revealed that subdiffusion yields a destabilisation of the basic state with respect to the standard diffusion $(r=1)$. The basic state turns out to be unstable for every positive Rayleigh number, $Ra > 0$.
\item\label{it2} Superdiffusion is linearly stable for every $Ra > 0$, even if a transient growth of the perturbation amplitudes occurs at earlier times of the evolution. Such a transient growth may well drive the linear evolution of the amplitude into a nonlinear domain, thus biasing the linear predictions at later times. In fact, the amplitude peaks achieved in the transient growth become larger and larger as the Rayleigh number increases above its standard diffusion critical value, $Ra_c = 4 \pi^2$.
\item Features \ref{it1}) and \ref{it2}) revealed a strong sensitivity of the instability phenomenon to every, even slight, departure from the standard diffusion process, with either subdiffusion or superdiffusion characteristics.
\end{enumerate}

A fractional-derivative model of subdiffusion has been also employed for testing its predictions relative to the linear stability of the basic state. It has been noted that a new dimensionless governing parameter $\xi$ is needed in this fractional derivative model. Such a parameter modulates the advection term in the mass diffusion equation. A solution for the linear stability analysis has been determined by adopting the Caputo fractional derivative with respect to time. It has been shown that the time-dependent perturbation amplitudes do not evolve exponentially, as is usual for linear stability analyses, but according to the Mittag-Leffler function of order $r$, where the argument is proportional to $t^r$. The instability occurs when the Rayleigh number is larger than the critical value, $Ra_c = 4\pi^2/\xi$. Such a critical value is compatible with that deduced for the special case of standard mass diffusion processes. In fact, by definition, it turns out that $\xi=1$ when $r=1$. Experimental data allowing for the evaluation of $\xi$, to date unavailable in the literature, might clarify whether subdiffusion effects are destabilising according to the fractional derivative model, inasmuch as predicted by the time-dependent diffusivity model. From another viewpoint, the experimental investigation of the Rayleigh-B\'enard instability could be a method for an indirect determination of the dimensionless parameter $\xi$ for subdiffusion processes.


\end{document}